\documentclass[%
 reprint,
 amsmath,amssymb,
 aps,nofootinbib,
]{revtex4-1}
\usepackage{bm}% bold math
\PassOptionsToPackage{linktocpage}{hyperref}
\usepackage[hyperindex,breaklinks]{hyperref}
\usepackage{enumitem}
\usepackage{slashed}

\renewcommand{\theta}{\vartheta}

\newcommand{\bra}[1]{\ensuremath{\left< #1\,\right|}}
\newcommand{\ket}[1]{\ensuremath{\left|\, #1\right>}}

\usepackage{array}
\usepackage{mathtools}

\usepackage{etoolbox}

\begin{document} 

\title{Black Holes as Brains: Neural Networks with Area  Law Entropy}

\author{Gia Dvali} 
%$^{1,2,3}$}
%\email{Georgi.Dvali@physik.uni-muenchen.de}
\affiliation{%
%$^1$ 
Arnold Sommerfeld Center, Ludwig-Maximilians-Universit\"at, Theresienstra{\ss}e 37, 80333 M\"unchen, Germany, 
}%
 \affiliation{%
%$^2$ 
Max-Planck-Institut f\"ur Physik, F\"ohringer Ring 6, 80805 M\"unchen, Germany
}%
 \affiliation{%
%$^3$ 
Center for Cosmology and Particle Physics, Department of Physics, New York University, 726 Broadway, New York, NY 10003, USA
}%

\date{\today}

\begin{abstract}
Motivated by the potential similarities between the underlying 
mechanisms of the enhanced memory storage capacity in black holes and in brain networks, 
we construct 
 an artificial quantum neural network based on gravity-like synaptic connections and a symmetry structure that allows 
to describe the network in terms of geometry of a $d$-dimensional space.  We show that the network possesses a critical state in which the gapless neurons emerge 
 that appear to inhabit a $d-1$-dimensional surface,
 with their number given by the surface area.  
 In the excitations of these neurons, the network  can store and retrieve an exponentially large number of patterns within an 
 arbitrarily narrow energy gap.  The corresponding micro-state entropy of the brain network exhibits an area law.    
The neural network can be described in terms of a quantum field, via identifying the different neurons with the different 
momentum modes of the field, while identifying the synaptic connections among the neurons with the interactions among the corresponding momentum modes.  Such a mapping allows to 
attribute a well-defined  sense of geometry to an intrinsically non-local system, such as the neural network,  and {\it vice versa}, it allows 
to represent the quantum field model as a neural network.  
\end{abstract}

%\pacs{(old) 14.60.Pq,13.15.+g,04.60.-m,11.30.Rd}

\maketitle

\section{Introduction} 
Given that black holes and human brains exhibit 
extraordinary memory storage capacities, it is natural to ask 
whether these systems share some very basic mechanism  
of the enhanced capacity of the information storage \cite{Gia}. 
Needless to say, a potential identification of such similarities, would  be mutually beneficial for  both fields of research.  Obviously, the final answer 
requires the involvement of biology, which is beyond our expertise. 
However, we can attempt to do a physicist part of the job. 
 A step in this direction was undertaken in \cite{Gia}, where 
 a general mechanism \cite{DG} - originally 
 proposed for the explanation of  an enhanced memory capacity in black holes - was applied to the framework \cite{kak}  of quantum neural networks and was shown 
 that it leads to an exponential enhancement of pattern storage and recognition capacities.  The idea is that by allowing the excitatory synaptic connections among the set of neurons to be {\it gravity-like}   - in the sense of a negative interaction energy -  the network 
 gives rise to a critical state with an exponentially enhanced  memory storage capacity.  This state is achieved due to the following effect. 
 A high excitation level of some neurons - due to their negative synaptic connections with others - lowers to zero the excitation threshold energies for the latter neurons, thereby making is possible to store an exponentially 
 large number of patterns within an infinitesimally  narrow energy gap.  Moreover, the patterns stored in such states can be redialed under an influence 
 of arbitrarily soft external stimuli. \\
 
 The  general message is that for gravity-like synaptic connections 
 in a quantum neural network the information storage mechanisms  can become potentially richer  than in more conventional approaches. 
Instead of storing patterns in the local minima of the energy 
function, as, for example, in  usual treatments of the Hopfield model \cite{Hopfield},  the enhanced memory points can be  identified among the highly 
excited critical states, where set of neurons becomes gapless. \\

   In this paper we shall attempt to take  one more step and try to
  see if a  neural network can possess some additional informational characteristics that are usually  considered to be the defining properties  of black holes, such as the Bekenstein's area law entropy \cite{Bek}.  \\
  
  In constructing such a network, we shall use the synergy of 
  the ideas displayed in \cite{Gia} and \cite{AREA}. 
 In the latter work it was shown that a theory of a quantum field inhabiting a 
 $d$-dimensional sphere and experiencing an 
 angular-momentum-dependent attractive interaction 
 exhibits a sharp enhancement of memory storage capacity 
 in a critical state in which the lowest angular momentum mode 
 is macroscopically excited.  This  excitation lowers the energy thresholds  
 of the higher angular momentum modes, so that the theory delivers a set of the gapless modes with their number 
 scaling as the area of a $d-1$-dimensional sphere. 
 The micro-states of these modes house an exponentially large number of patterns, with the corresponding entropy given by the area of the same $d-1$-dimensional sphere. \\

 The quantum brain network that we shall construct in the present paper is nothing but a neural network implementation of the same model.  This is achieved by means of identification of the neuron-describing degrees of freedom  with the number operators of the angular harmonics of the quantum field.  Such mapping provides a way in which  the  sense of locality and geometry can be brought into something intrinsically non-local, such as the neural network.  \\

 However, in our construction we shall not simply copy the quantum field 
model,   but rather arrive to the desired neural network 
 by a bottom-up approach. 
   For this purpose we design a neural network in which the synaptic connections are closer cousins to the gravitational 
  interaction  as compared to 
   the ones studied in \cite{Gia}. Namely, we study a network 
in which the synaptic connections $\hat{W}_{kr}$ among the pairs of neurons $(k,r)$ are energetically negative and also proportional to the
products of their energy thresholds. \\
 
 Again, as in the case of \cite{Gia}, we observe that such a  network exhibits a critical state in which a 
set of would-be-high-threshold neurons becomes effectively  
gapless. Correspondingly, in this set of neurons the network can store an exponentially large number of patterns
within an arbitrarily narrow energy gap.  \\

The entropy of the resulting micro-states depends on the
symmetry structure of the network. We show that, when 
we impose a symmetry group of a $d$-dimensional sphere, a set of the gapless neurons emerges with their number equal to an area of a $d-1$-dimensional sphere  of the same radius.  Correspondingly, the micro-state entropy of the patterns encoded in the excitations of the emergent gapless neurons
also scales as the area of the same $d-1$-dimensional sphere. \\

 Thus, it appears that  the neural network - when translated in quantum field theoretic language - 
  can exhibit some intrinsic {\it holographic}  properties, such as, the 
 emergence of the gapless neurons inhabiting a lower dimensional surface with the  resulting micro-state entropy that scales as the corresponding area. This is something highly reminiscent of gravitational systems, such as, the black holes \cite{Hologram} or the AdS space \cite{ADS1,ADS2,ADS3}. \\

 \section{dictionary} 
 
 In order to import certain ideas from black holes to the neural networks, 
 we need to establish a dictionary. 
  First, we need to connect the basic degrees of freedom in the two systems. 
 Let us first ask, what are the black holes from the quantum field theoretic point of view? 
 
 Although a fully-satisfactory quantum description of black holes is still missing, it is reasonably clear      
  that they represent the states in a Hilbert space of  quntized fields.   
   We shall not question 
  this statement.  Our task thus boils down to  the need of establishing a dictionary between the states 
  of the qiantum fields and the states of the quantum neural networks.    
 In quantum field framework  the elementary degrees of freedom are the quantum oscillators describing the different momentum (and spin/helicity) modes of the field.  
   On the other hand, in a neural network the elementary degrees of freedom are the excitations of neurons.   We shall represent both of these entities by the  
creation and annihilation operators 
$\hat{a}_k^{\dagger}, \hat{a}_k$, that satisfy the usual    
 algebra, 
       \begin{equation} 
    [\hat{a}_r,\hat{a}_k^{\dagger}] = \delta_{rk}\,, \, \, 
  [\hat{a}_r,\hat{a}_k]  =   [\hat{a}_r^{\dagger},\hat{a}_k^{\dagger}] =0\,,   
    \label{algebra} 
 \end{equation} 
 where $k$ and $r$ are the labels. 
  In case of a quantum field  they  identify the various momentum (or angular-momentum) modes, whereas in a neural network they represent the labels of the neurons.   
 The excitation level of a neuron $k$  in a given quantum state 
   is described by an eigenvalue (or an expectation value)  of the corresponding number operator $\hat{n}_k \equiv \hat{a}_k^{\dagger}\hat{a}_k$.
  Thus, the basic states of a neural network are the 
  Fock states $\ket{n_1,n_2,.....}$ labeled by the excitations levels 
  of different neurons.  Since on the quantum field theory side the analogous number operators describe the occupation numbers of respective momentum modes, the excitation level of a neuron
 is translated as the occupation number of a particle mode. \\
 
  Next thing is the interaction. The quantum neurons influence each other through the synaptic connections, which can be represented as an 
operator with two or more indexes,  $\hat{W}_{kr}$.  When it is energetically favorable 
 for the two neurons $k$ and $r$ to simultaneously be in the excited   
 states, we shall call the synaptic connection {\it excitatory} in the {\it energetic sense}.  Correspondingly, in the opposite case 
 we refer to the connection as {\it inhibitory}.   For the quantum field 
 this translates as the interaction energy between the two modes 
 $k$ and $j$ being negative (attractive) or positive (repulsive).  \\
 
 Now, the gravitational interaction among the particle modes is attractive 
 and its strength is set by their energies. 
 Thus, in order to imitate gravity-like connections among the 
 neurons, we shall make the synaptic connection among the 
 two neurons $k$ and $r$ negative (excitatory) and moreover proportional to the product of their threshold energies
\begin{equation} 
 \hat{W}_{kr} \propto  - \epsilon_k\epsilon_r\,.    
\label{product} 
\end{equation} 
  
Next, we need to come up with some rough guideline telling us  for what type of states we should be looking for in the neural network. 
 This guideline is very simple. From, the quantum field theoretic perspective defined on Minkowski space, 
 The Fock vacuum corresponds to a zero occupation number of all the modes, 
 $\ket{vac} = \ket{0,0,.....}$.   The black hole state is certainly not a Fock vacuum, but rather some highly excited state. Indeed, to form a macroscopic black hole of  certain size $R$, we need to produce a very high occupation number of the low momentum quanta, localized 
 within that region of space.  In general, even in a pure gravity theory,  a macroscopic black hole 
 effectively consists of many quanta due to a high occupation number of the constituent soft gravitons that form 
its gravitational field \cite{GCNNN}.

 For our purposes, this very crude guideline will be enough.  This guideline is telling us that
 with the black hole analogy, we need to look for a state of high memory storage capacity among the states in which 
 some of the low-threshold neurons
 are highly excited, i.e., corresponding quantum oscillators have the high occupation numbers. 
 \\
 
  To summirize, we establish the following dictionary: 
 
 \begin{itemize}
  \item Particle momentum mode  $\leftarrow \rightarrow$   Neural degree of freedom;  
 \item  Ocillator energy gap  $\leftarrow \rightarrow$ Excitation energy threshold of neuron;   
  \item Mode occupation number $\leftarrow \rightarrow$ Excitation level of neuron; 
  \item Gravitational interaction among particles  $\leftarrow \rightarrow$ Excitatory synaptic connection among neurons; 
  
 \item Black hole state $\leftarrow \rightarrow$ Critical state of highly excited low threshold neurons.  
 
\end{itemize} 

 The above dictionary gives us a rough idea that in a neural network designed 
by the above rules,  we should encounter a state 
of  an enhanced memory storage capacity in form of a state 
in which some of the lowest threshold neurons are highly excited.    
 As we shall see, this is indeed the case. \\

  Finally, we would like to comment that our main goal is to reach 
 a maximal energy efficiency of the memory storage, as well as, the efficiency of the response to the input external patterns.  We shall not be interested in implementation of the learning algorithms and other computational tasks in the quantum neural networks.

\section{Network} 

 Consider a quantum neural network where  
neurons are labelled by a set of integers $k_1,...k_d$, which 
we shall briefly  denote by $k$.  Each neuron corresponds to a quantum 
degree of freedom represented by the creation and annihilation 
 operators $\hat{a}_k^{\dagger}, \hat{a}_k$. 
 They satisfy the commutation relations (\ref{algebra}),  
 where $k$ and $r$ denote the two sets $k_1,...k_d$ and 
 $r_1,...r_d$ respectively, and $ \delta_{rk} \equiv \delta_{r_1k_1}\delta_{r_2k_2}... \delta_{r_dk_d}$. \\

The Hamiltonian of the network has the following form, 
\begin{equation}
\label{HA} 
 \hat{H} = \sum_k \epsilon_k \hat{a}_k^{\dagger}\hat{a}_k - 
 \sum_{k,r} \hat{a}_k^{\dagger} \hat{W}_{kr} \hat{a}_r   
\end{equation}
 where the summation is taken over all sets. 
 Here $\epsilon_k$ is the threshold energy required for the excitation of the neuron $k$.  An each level $\epsilon_k$ exhibits a certain degeneracy ${\mathcal N}_k$ due to a 
symmetry structure of the network that we shall discuss below. 
The Hamiltonian (\ref{HA}) can be viewed as a generalized quantum version 
of the Hopfield model  \cite{Hopfield}, but with some important
modifications that we shall now specify.   \\

 The operator $\hat{W}_{kj}$ is a synaptic matrix operator. The idea is to choose it to be {\it gravity-like}:  
\begin{equation} \label{WW} 
 \hat{W}_{kr}  \equiv 
    {\epsilon_k  \epsilon_r \over 2\Lambda} \left(3 - 
    {\epsilon_k  \epsilon_r \over \epsilon_*^2} \right )
     \sum_{s,q} C_{skqr}  \hat{a}_s^{\dagger} \hat{a}_q\,, 
 \end{equation}
  where $\Lambda$ is a parameter of dimensionality of energy, whereas $C_{skqr}$ are dimensionless coefficients that determine the structure of the synaptic connections. They will be specified later.    
  The relative factor $3$ in the brackets is introduced for convenience. \\ 
 
  The parameter $\epsilon_*$ determines the critical level $k_*$, which represents the turning  point.  
  From (\ref{WW}) it is clear that the synaptic connection 
  energy among the neurons changes sign at 
   $\epsilon_{k}\epsilon_{r} = 3 \epsilon_*^2$. 
  The over-all sign will be normalized in such a way that the 
  connection energy is negative for  $\epsilon_{k}\epsilon_{r} <  3\epsilon_*^2$.
  Thus, in this regime, the connections are  {\it excitatory in the energetic sense}. That is,  an excitation of one nauron makes it energetically favourable to excite others.  
  In other words, an excitation of a given neuron, lowers the excitation thresholds of all the other neurons connected to the former neuron. 
  
\begin{figure}
 	\begin{center}
        \includegraphics[width=0.53\textwidth]{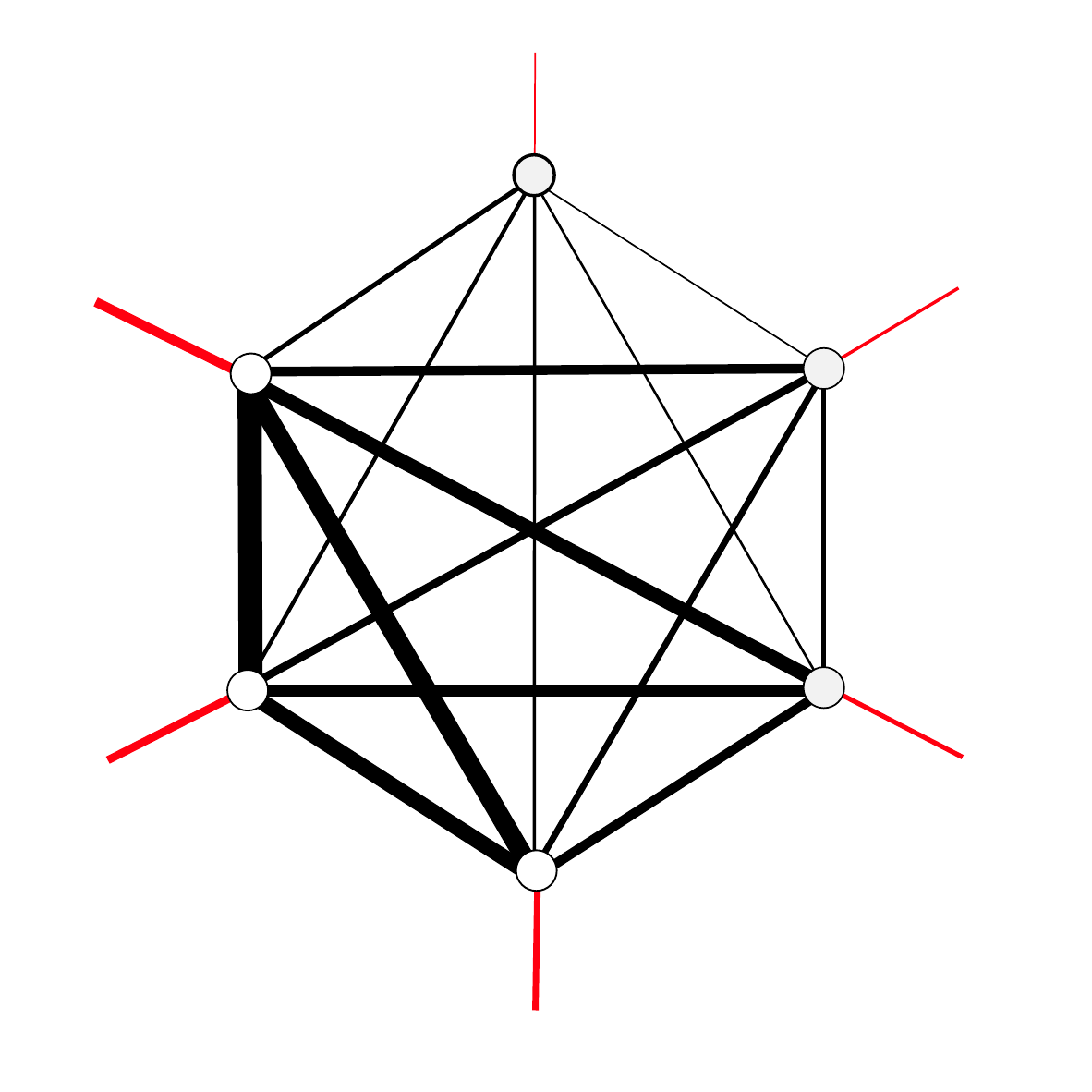}
 		\caption{The neural network example for six neurons and the choice $C_{skqr}  = 
 \begin{cases}
     \delta_{sr}\delta_{kq} & \text{for} ~ k\neq r, \\
     0 & \text{otherwise}
\end{cases} $.  For definiteness, the regime  $\epsilon_{k}\epsilon_{r} \ll \epsilon_*^2$ 
is assumed. 
 The neurons are represented by white circles. 
The red lines represent the inhibitory biases due to the positive threshold energy gaps, whereas the black lines represent 
the excitatory connections due to the energetically negative couplings. The strength of the excitatory connection is proportional to the product of the corresponding threshold energies and is expressed through the thickness of the corresponding synaptic lines. } 		  		
\label{EXnet}
 	\end{center}
 \end{figure}

 Correspondingly, in the opposite regime, $\epsilon_{k}\epsilon_{r} >  3\epsilon_*^2$, the connection energy becomes positive. 
 This fact guarantees that the energy of the network is bounded from below. 
  Indeed, for $\epsilon_{k}\epsilon_{r} > 3\epsilon_*^2$ the synaptic connection becomes inhibitory, thereby preventing the network from instability.  
   An example of a neural network for a particular choice of 
 $C_{skqr}$ is depicted on Fig.\ref{EXnet}. 
       \\
  
   Notice, the above synaptic connection among the neurons, in the excitatory regime, is analogous to a gravitational interaction among particles.  Therefore,
  such a setup represents a most naive way of 
   bringing a {\it gravity-like} dynamics into the neural networks. 
   However, as we shall see, despite its simplicity,  it suffices for  generating an exponential enhancement of memory storage and pattern recognition capacities with the corresponding entropy that obeys an area law. \\
   
   We shall now impose a spherical symmetry on the structure of the 
neural network.  We shall do this in the sense of a Fourier-transform, by assuming 
$C_{skqr}  \equiv \int d^d\Omega Y_{s}^* Y_{k}^*Y_{q}Y_{r}$,
where,  $Y_k(\theta_a)$ are the harmonic functions on
a $d$-dimensional unit sphere, with the set of angular coordinates
$\theta_a,~a=1,...d$,  a volume element 
$d^d\Omega$ and a total volume $\Omega$.  
Correspondingly, the label $k$ stands for a set of $d$ integers, i.e., 
$Y_k \equiv Y_{k_1,...,k_d}$, which satisfy,   
 $|k_1|\leqslant k_2 \leqslant ...\leqslant k_d = 0,1,...,\infty$. 
 These spherical harmonics form a complete orthonormal set. In particular: 
 $\int d^d\Omega  \, Y_k^*Y_{r} = \delta_{kr}$.
  The functions $Y_k$ represent the eigenfunctions of the covariant  Laplace operator on the $d$-sphere of radius $R$, 
   \begin{equation} 
 \Delta  Y_k\,   = - {k_d(k_d + d-1) \over R^2} Y_{k} \, , 
  \label{eigen} 
  \end{equation} 
where each eigenvalue $k_d(k_d + d-1)$ exhibits the following  degeneracy: 
$ {\mathcal N}_k =  \sum_{k_{d-1}=0}^{k_{d} }  \sum_{k_{d-2}=0}^{k_{d-1}}...\sum^{k_2}_{k_{1} = - k_2} \sim  (k_d)^{d-1}$.  Thus, in this description the neural 
modes $\hat{a}_k^{\dagger}, \hat{a}_k$, represent the creation 
and annihilation operators for the angular harmonics corresponding to 
a given set $k$. The correspondence between the threshold 
energies $\epsilon_k$ and the angular-momentum levels is enforces by the spherical symmetry through the relation: 
   \begin{equation} \label{dictionary} 
   \epsilon_k \equiv  \kappa {k_d(k_d + d-1) \over R^2}\, 
  \end{equation} 
  where $\kappa$ is a parameter of the network, with 
  the dimensionality $[\kappa] = [energy] \times [distance]^2$. 
  For the compactness of notations, we shall  absorb it into the redefinition of 
  $\Delta$.  
 \\

%%%%%%%%         
   
    The above setup allows us to identify the different neurons  
  with the corresponding angular momentum modes of a  non-relativistic quantum field
  $\hat{\psi}(\theta_a)$ living on a $d$-dimensional sphere. 
  That is,  
  \begin{equation} 
      \hat{\psi} = \sum_{k} \, Y_k(\theta_a) \hat{a}_k \,. 
     \label{expansion} 
   \end{equation}   
 The description of the neural network in terms of Fourier components 
 of a local field, allows to give a geometric interpretation to something intrinsically non-local. Indeed, neurons interact non-locally:  Everybody talks to each other according to the rules determined by the synaptic connections, with no real sense of local geometry. \\
 
 However,  the sense of geometry is provided by the concept of a quantum field.
  Indeed, while the interaction of the quantum fields in position space is local, in momentum space it is not.   The fact that the neurons are translated into the momentum modes, shows how to 
 connect them to geometry.  \\

In terms of a quantum field the Hamiltonian of the network
(\ref{HA}) can be written in the following form: 
  \begin{eqnarray} \label{H1}  
   && \hat{H}  =   \int d^d\Omega \,\, \Big[-  \hat{\psi}^{\dagger} \Delta  \hat{\psi}\,  -  \, {3 \over 2\Lambda}  \,  (\hat{\psi}^{\dagger}  \Delta  \hat{\psi}^{\dagger}) (\hat{\psi} \Delta  \hat{\psi})\, + \nonumber \\
   && + \, {1 \over 2\Lambda\epsilon_*^2}  \,  (\hat{\psi}^{\dagger}  \Delta^2  \hat{\psi}^{\dagger}) (\hat{\psi} \Delta^2  \hat{\psi}) \Big]\,.
   \end{eqnarray} 
 This Hamiltonian represents the model of \cite{AREA}, which describes a non-relativistic bosonic quantum field living on a  $d$-dimensional sphere and experiencing an 
 angular momentum dependent attractive interaction. 
 
 As shown in \cite{AREA}, this system possesses a critical state 
 with exponentially enhanced memory capacity and the micro-state 
 entropy that scales as the area of a lower dimensional sphere, $S_{d-1}$, of the same radius. \\
 
 Not surprisingly, the above property translates into an analogous property of the neural network, once we  map the angular harmonics onto neuron degrees of freedom. 
   In order to see this, we can simply repeat the analysis 
 of  \cite{AREA}, which is ready-made for our purposes. \\

 The critical state of an enhanced memory capacity 
 is reached due to the high excitations
 (i.e., the high occupation numbers) of some of the neurons, which - due to 
 the excitatory synaptic connections - lowers  the excitation thresholds for others.    We choose the highly excited neuron to be 
 the one with the lowest energy threshold. This corresponds to   $k=0$ harmonics.   \\

 We thus focus our attention on the states in which only 
 the $k=0$ neuron is highly excited,  with the excitation level  given by  a large occupation number of the zero mode 
 $\langle \hat{a}_0^{\dagger} \hat{a}_0 \rangle  = N_0 \gg 1$.  
 
 In order to derive the effective Hamiltonian on such a state, we can effectively replace   
 the operators  $\hat{a}_0^{\dagger},  \hat{a}_0$ by the 
 $c$-numbers,  $\hat{a}_0^{\dagger} = a_0 \,,~
 \hat{a}_0 = a_0^*$, with  
 $|a_0|^2=N_0 $.  That is, we are using the 
 Bogoliubov approximation \cite{bogoliubov}  for the quantum neural network. 
  The existence of a critical state with an increased 
  memory capacity can be cleanly demonstrated in a 
  double-scaling  limit.  Following \cite{AREA}, we take   
   \begin{equation} 
 N_0 \rightarrow \infty\,, ~ ~  \Lambda \rightarrow \infty \,,  
 ~~ {N_0 \over \Lambda}  = {\Omega \over \epsilon_*} ={\rm finite} \,.
 \label{doublescale}
 \end{equation} 
In this limit in (\ref{WW}) only the terms with $s=q=0$ and $k=r\neq 0$ survive, 
because of the following reasons. 
First notice,  the terms with either  $k$ or $r$ zero, vanish due to accompanying $\epsilon_k$ and $\epsilon_r$ coefficients.
Next, the terms with either $s$ or $q$ non-zero scale as 
${1 \over \sqrt{N_0}}$, whereas the terms with both $s$ and $q$ nonzero scale as  ${1 \over N_0}$.
Finally, we use the property $C_{0k0r} = {\delta_{kr} \over \Omega}$.

Thus, in the double scaling limit the Hamiltonian takes the following form: 
  \begin{equation}
 \hat{H}  =    \sum_{k} \omega_k 
  \hat{a}_k^{\dagger}\hat{a}_k \,   + \, {\mathcal O}\left({1\over \sqrt{N_0}}\right )\,,
 \label{HAL} 
 \end{equation}
where, 
 \begin{equation}
\omega_k \equiv {\epsilon_k \over 2} \left(
 {\epsilon_k \over \epsilon_*} - 1\right)^2 
 \left( {\epsilon_k \over \epsilon_*} + 2\right)\,. 
\label{GAP}
\end{equation}
 Thus,  we arrive to a one-parameter family of neural networks
 labeled by the critical level $\epsilon_*$.   The Hamiltonian 
is semi-positive-definite  and touches zero only for 
$\epsilon_k=0$ and $\epsilon_k = \epsilon_*$.  The first point simply 
reflects the gaplessness of the zero angular-momentum mode and  
exhibits no interesting degeneracy.  The second point is the most  interesting. 
This point shows that the threshold energies  for  all the neurons with $k \in k_*$ collapse to zero.  That is, they become exactly {\it gapless}. 
  
  The number of gapless neurons is equal to the degeneracy of the eigenvalue  $ \epsilon_k = \epsilon_*$, which is given by  $N_{gapless} = {\mathcal N}_{k_*} \sim  k_*^{d-1}$.
 This quantity  scales as an area of a $d-1$-dimensional sphere: 
   \begin{equation} 
      N_{gapless} \, = \, \left ( {R \over L_*} \right )^{d-1} \, . 
   \label{Nmodes}
   \end{equation}  
  where $L_* \equiv \sqrt{\kappa  \over \epsilon_*}$. 
  It is convenient to relabel the gapless neurons $k \in k_*$
 by an additional index $\gamma$ as  $\hat{a}_{k_*\gamma}^{\dagger} \,, \hat{a}_{k_*\gamma}$, where the new index takes the values $\gamma = 1,2,...,N_{gapless}$.  The corresponding number operators 
shall be denoted by $\hat{n}_{*\gamma} \equiv \hat{a}_{k_*\gamma}^{\dagger}\hat{a}_{k_*\gamma}$. \\    
    
   In the same time, as it is clear from (\ref{GAP}),  all neurons  with $k \neq k_*$   (except $k =0$) have positive and large energy gaps
  $\omega_k$.  Therefore, the storage of information in such neurons is very costly in energy. Because of 
  this reason, they shall not be counted in the estimate of the memory storage capacity of the system. 
   
 \section{Micro-States}   
   
    The patterns can be stored in the set of the basic states, in which the  
excitation levels of the $k \in k_*$-neurons assume different possible values on an interval, 
$0 \leqslant\langle \hat{n}_{*\gamma} \rangle  < n$,
where $n$ is some maximal value. 
 The basis can be  conveniently  chosen, for example, in form of the number-eigenstates or the coherent states of the gapless modes.  \\

The number eigenstate basis is represented by the Fock space 
 ket-vectors of the form,   
\begin{equation} \label{nstate} 
\ket{n_{*1} , n_{*2},...n_{*N_{gapless}}} 
\end{equation} 
 where, $0 \leqslant  n_{*\gamma}  < n$ are the eigenvalues of
  $ \hat{n}_{*\gamma}$-operators.

 The number of such micro-states is equal to 
${\mathcal N}_{states} = n^{N_{gapless}}$, with the corresponding micro-state entropy given by, 
  \begin{equation} 
      {\rm Entropy}  \, \sim \, \left ( {R \over L_*} \right )^{d-1} ln(n) \, . 
   \label{ENTR}
   \end{equation}  
  Notice, in the double-scaling limit (\ref{doublescale}),  $n$ can be taken arbitrarily large at no cost in  energy, because 
 in this limit the interaction terms vanish and $k \in k_*$ modes are exactly gapless.
   However, in reality, the value of $n$ will be restricted by the following effects. 
   \begin{enumerate}
  \item   First, irrespective of the value of $N_0$, the value of $n$ will be constrained   by the resolution capacity of a reading device that retrieves and decodes the stored patterns from the memory of the network. 

  \item Secondly, $n$ will become restricted by the finite-$N_0$-effects as soon as we move away from the double-scaling limit. 
   
\end{enumerate}

Notice, since the Hamiltonian  (\ref{H1}) conserves the total particle number, 
 the micro-states are split into the super-selection sectors according to this number.  
   The degenerate micro-states are obtained by the small redistributions of the total occupation number $N_0$ among the zero mode, $k=0$, and the rest of the gapless modes, $k \in k_*$. 
 That is, in the states given by (\ref{nstate}), the occupation number of the zero mode is replaced by,  $N_0 \rightarrow N_0 - \sum_{\gamma=1}^{N_{gapless}} n_{*\gamma}$,
in order to keep the total particle number intact.\\
      
   In the double-scaling limit (\ref{doublescale}), since $N_0=\infty$, 
  such a re-adjustment of $N_0$  causes no back-reaction on the energy gaps,  for any finite value of  $n$.  
   Correspondingly, there is no restriction on $n$, and the number 
   of micro-states diverges as  $log(n)$ for large $n$.  
   However, for a finite $N_0$, such a redistribution will lead to the small-but-finite  level-splittings
   among the different micro-states.  This will restrict the value of $n$ from above.   
   However,  this level-splittings will be suppressed by powers of $1/N_0$ and can be made arbitrarily 
   small by taking large enough $N_0$. 
   
      Also,  for any large-but-finite number $N_0$, there will be a maximal number $n$ beyond which the interaction terms  
shall become important and shall start costing a high energy.  The good news is that all the above effects can be kept under control at large $N_0$. \\

 In addition, due to the conservation of the angular momentum, the micro-states split into the super-selection sectors corresponding to the different values 
 of the total angular momentum. \\ 
 
   Finally, let us comment that an alternative basis for the information storage  can be formed by the coherent  states of the  gapless modes, instead of their number eigenstates. Such states, $\ket{\alpha_{*1},..., \alpha_{*N_{gapless}}}$, are the eigenstates of the destruction operators $\hat{a}_{*\gamma}$ and shall be labeled by the set of the corresponding  
 complex  eigenvalues $\alpha_{*\gamma}$, 
  \begin{equation} \label{coherent}
  \hat{a}_{*\gamma}\ket{\alpha_{*1},..., \alpha_{*N_{gapless}}} = 
 \alpha_{*\gamma} \ket{\alpha_{*1},..., \alpha_{*N_{gapless}}} \, .
  \end{equation} 
  Here $\alpha_{*\gamma}$ are the complex parameters that among other 
  properties, satisfy the obvious relation $|\alpha_{*\gamma}|^2 = \langle \hat{n}_{*\gamma} 
 \rangle$.
 The two coherent states are almost orthogonal provided they satisfy 
 $ \sum_{\gamma}|\alpha_{*\gamma} - \alpha_{*\gamma}'|^2 \gg 1$.
Such a set of the coherent states is approximately classical,
provided the mean occupation numbers are large, i.e., $|\alpha_{*\gamma}| \gg 1$.  The 
storage of information in the coherent states is especially useful for generalizing 
the phenomenon of the critical enhancement of memory to the classical 
brain networks, as discussed in \cite{Gia}. In such a case, from the information storage point of view, the neural network becomes equivalent  
to the set of classical harmonic oscillators of very low frequencies, with the patterns being stored in their amplitudes and the phases.

   \section{Response to External Patterns} 
   
   Not surprisingly, the neural network in the above critical state 
  can respond - with maximal precision - to an arbitrarily soft  
 input stimuli.   We can briefly repeat the analysis of \cite{Gia}
 for the present network.  
  For this purpose, we shall introduce a new layer of neurons with corresponding  
 creation/annihilation operators,  $\hat{b_k}^{\dagger}, \hat{b_k}$, 
satisfying  the usual commutation relations, 
  $  [\hat{b_r},\hat{b_k}^{\dagger}] = \delta_{rk}\,, \, \, 
  [\hat{b_r},\hat{b_k}]  =   [\hat{b_r}^{\dagger},\hat{b_k}^{\dagger}] =0,$    
 and commuting with $\hat{a_k}^{\dagger},\hat{a_k}$-operators.  
 We shall assume that a pattern encoded in an initial state 
 of $b$-neurons is transmitted to $a$-neurons through 
 the synaptic connections.
 The response efficiency will be measured 
by its precision, i.e., by how accurately the state of $a$-neurons can reproduce an input pattern. Thus, $b$ and $a$ modes play the 
roles of the input and the output  layer neurons respectively. 
  We are specifically interested in the {\it energy efficiency}  of the response, i.e., a situation in which the output layer 
  $a$ is capable of accurately responding to the external patterns of a very low energy.  For simplicity, we shall couple the input and the output layer neurons via the diagonal synaptic connections. \\

  If we represents the 
 $b$-neurons as the angular momentum modes of a quantum field  
  \begin{equation} 
     \hat{\chi} = \sum_{k} \, Y_k(\theta_a) \hat{b}_k \,, 
    \label{expansion} 
    \end{equation}
 the coupling between the $a$ and $b$ neurons can be written 
 in a compact way as
    \begin{equation} 
    \hat{H}_{ab}  =  - \int d^d\Omega \, q (\hat{\psi}^{\dagger} \hat{\chi}\,  + \hat{\chi}^{\dagger} \hat{\psi})\, 
    +   \kappa' \hat{\chi}^{\dagger} \Delta  \hat{\chi}\,, 
 \label{HCHI} 
 \end{equation}  
 where $q$ and $ \kappa'$ are parameters. 
 Rewriting in momentum space, we get,   
    \begin{equation} 
    \hat{H}_{ab}  =  q \sum_{k \in k_*}  (\hat{b}_k^{\dagger} \hat{a}_k  + \hat{a}_k^{\dagger} \hat{b}_k )  +  \epsilon_b \sum_{k \in k_*} \hat{b}_k^{\dagger}\hat{b}_k \, + \, ...,
 \label{HX} 
 \end{equation}  
 where,  $\epsilon_b$ measures the energy gap for the minimal excitation level of the input neurons, and $q$ parameterizes the strength of the synaptic connection between the input and the output neurons.
 Since we wish the input stimulus to be very soft, we shall take 
 both $\epsilon_b$ and $q$ to be very small. 
   
 Notice, because the $a$-neurons with  $k \notin k_*$ have very high energy thresholds, they play no role in recognition of patterns fed by the soft external stimuli carried by $b$-neurons.  Therefore,
  $k \notin k_*$ modes of $b$-neurons were disregarded in (\ref{HX}). So the sum only runs over the modes belonging to the critical level $k\in k_*$ and over the corresponding modes from the $b$-sector.   For convenience, we shall label the latter $b$-modes 
 by the same index $\gamma$ as the gapless $a$-modes: $\hat{b}_{*\gamma} \in  \hat{b}_{k_*}$, where $ \gamma =1,2,...N_{gapless}$.  
 \\
 
  The interaction Hamiltonian (\ref{HX}) ensures that a pattern - initially encoded in the state of $b$-neurons - will get transmitted to 
 $a$-neurons.   
The accuracy of the response is determined by how precisely 
the output state of $a$-neurons copies  the input pattern 
of $b$-neurons.  In the same time, the energy cost-efficiency of recognition is measured by the lower bound on the softness of the recognized input pattern. \\

 Intuitively it is clear that the critical state must exhibit a maximal energy-cost-efficiency 
 of recognition, since the gapless $a$-modes are able to align to an external pattern at an arbitrarily low energy cost.  Indeed, let the initial input pattern
 be given by some {\it coherent} state of the $b$-modes 
 $\ket{in}_b \equiv \ket{\beta_{1},... \beta_{N_{gapless}}}$, i.e., 
 $\hat{b}_{*\gamma} \ket{in}_b = \beta_{\gamma}  \ket{in}_b$, where 
 $\beta_{\gamma}$ are the complex eigenvalues. 
 Thus, the  original pattern is encoded in form of the 
 expectation values of the occupation numbers of 
  $b$-neurons, with the pattern vector being $(|\beta_1|^2, |\beta_2|^2, ...,|\beta_{N_{gapless}}|^2)$.    
Let the initial state 
 of gapless $\hat{a}_{*\gamma} $-neurons be given by their Fock vacuum $\ket{in}_a = \ket{0,0,0,...0}$ of the critical state.  \\
 
 The time-evolution of the state vector of the entire system
 is given by  $\ket{t} = e^{-{i\over \hbar} \hat{H}_{ab}t }\ket{in}_b\otimes\ket{in}_a$, where we took into the account the double-scaling limit and the fact that the rest of the Hamiltonian stays 
  zero throughout the evolution. 
      
It is easy to obtain the time-evolution of the expectation values of the excitation levels of $a$-neurons. For example, 
for $\epsilon_b=0$, we have \cite{giamischa},\cite{Gia}: 
  \begin{equation} 
\bra{t} \hat{a}_{*\gamma}^{\dagger} \hat{a}_{*\gamma} \ket{t} \, = |\beta_{\gamma}|^2 {\rm sin}^2\left ({tq \over 2\hbar}\right ) \, , 
  \label{evol1} 
   \end{equation}
  Thus, after time $t={\pi \hbar\over q}$,  the state of output 
 $a$-neurons,  copies the input pattern of $b$-neurons. 
 In other words, the information can be encoded (or retrieved) 
 at arbitrarily small energy cost. 
 Due to criticality, the neural network is sensitive to arbitrarily 
 soft external stimuli. \\ 
 
 \section{Discussions} 
 
  We have seen that a quantum brain network with the gravity-like 
  synaptic connections can acquire a very high complexity and an ability to store an exponentially large number of patters within a narrow energy gap.  Assuming a spherically symmetric structure of the synaptic connections, the entropy of the critical states obeys the area low.
  In such a state the network delivers a set of the gapless neurons that 
 inhabit a sphere of one dimension less.  
 The identification of neurons with the angular-momentum modes 
of a quantum field gives a well-defined geometric meaning to the neural network. \\
  
  The considered phenomenon of the critical memory enhancement has a smooth classical limit and can thereby take place is classical neural networks.  We have seen this explicitly by storing 
 patterns in the coherent states of the gapless modes 
 (\ref{coherent}).  In the regime in which their mean occupation numbers
 are large, the states are effectively classical. 
   The detailed discussion of classicality  goes exactly as 
 in \cite{Gia} and will not be repeated here.   \\
 
 The above features may serve as a supporting evidence that 
 some version of black hole information storage may be taking 
 place within classical brains. \\

  Finally, we wish to stress that by no means our result should be interpreted as an indication that the real quantum gravity effects play 
 any role in memory storage in biological brains. Such suggestions have been made in the past \cite{Nanopoulos} and have been partially addressed by \cite{tegmark}.  This is not the question we are trying to discuss. 
  \\
 
  Our idea is fundamentally different. Following \cite{Gia},
 we suggest that the enhanced memory  capacity mechanism  that operates in black holes is very general  in its essence  and goes beyond the gravitational systems. Moreover,  it possess a well-defined classical counterpart. Correspondingly, this mechanism can 
  be operative in the brain networks, both quantum and classical. 
  \\
   
   The key is the  phenomenon of criticality that takes place in systems with gravity-like synaptic connections among the 
bosonic degrees of freedom that can be in highly excited 
(highly occupied) states.  In such states many gapless modes emerge. For a certain symmetry structure of the network, the gapless modes appear to inhabit a lower dimensional surface area, thereby displaying a  holographic behaviour.

\section*{Acknowledgements}
 We thank Lasha Berezhiani, Allen Caldwell, Cesar Gomez, 
 Tamara Mikeladze-Dvali and Sebastian Zell for discussions. 
An exchange on \cite{AREA} with Oliver Janssen is acknowledged. 
   It is a pleasure to point out an ongoing communication with Dmitri Rusakov and Leonid Savchenko
   about realistic neural networks.  
This work was supported in part by the Humboldt Foundation under Humboldt Professorship Award, ERC Advanced Grant 339169 "Selfcompletion", by TR 33 "The Dark Universe", and by the DFG cluster of excellence "Origin and Structure of the Universe". 

\appendix

\end{document}